\documentclass[%
 reprint,
superscriptaddress,
 amsmath,amssymb,
 aps,
]{revtex4-2}

\usepackage{graphicx}
\usepackage{dcolumn}
\usepackage{bm}

\begin{document}

\preprint{APS/123-QED}

\title{Tracing non-Abelian anyons via impurity particles}

\author{Niccol\`o Baldelli}
\affiliation{ICFO-Institut de Ciencies Fotoniques, The Barcelona Institute of Science and Technology, 08860 Castelldefels (Barcelona), Spain}
\author{Bruno Juli\'{a}-D\'{i}az}%
\affiliation{Departament de F\'{i}sica Qu\`{a}ntica i Astrof\'{i}sica, Facultat de F\'{i}sica, Universitat de Barcelona, 08028 Barcelona, Spain.}
\affiliation{Institut de Ci\`{e}ncies del Cosmos, Universitat de Barcelona, ICCUB, Mart\'{i} i Franqu\`{e}s 1, Barcelona 08028, Spain}
\author{Utso Bhattacharya}
\affiliation{ICFO-Institut de Ciencies Fotoniques, The Barcelona Institute of Science and Technology, 08860 Castelldefels (Barcelona), Spain}
\affiliation{
Max-Planck-Institut für Quantenoptik, D-85748 Garching, Germany}
\author{Maciej Lewenstein}
\affiliation{ICFO-Institut de Ciencies Fotoniques, The Barcelona Institute of Science and Technology, 08860 Castelldefels (Barcelona), Spain}
\affiliation{ICREA, Pg. Lluis Companys 23, 08010 Barcelona, Spain}
\author{Tobias Gra{\ss}}
\affiliation{ICFO-Institut de Ciencies Fotoniques, The Barcelona Institute of Science and Technology, 08860 Castelldefels (Barcelona), Spain}

\date{\today}

\begin{abstract}
Non-Abelian excitations are an interesting feature of many fractional quantum Hall phases, including those phases described by the Moore-Read (or Pfaffian) wave function. However, the detection of the non-Abelian quasiparticles is challenging. Here, we consider a system described by the Moore-Read wave function, and assume that impurity particles bind to its quasiholes. Then, the angular momentum of the impurities, reflected also by the impurity density, provides a useful witness of the physics of the non-Abelian excitations. By demanding that the impurities are constrained to the lowest Landau level, we are able to write down the corresponding many-body wave function describing both the Moore-Read liquid and the impurities. Through Monte Carlo sampling we determine the impurity angular momentum, and we show that it suggests a quantum-statistical parameter $\alpha = a\nu -b +P/2$ for the quasiholes, where $\alpha$ ranges from $0$ for bosons to $1$ for fermions. A reasonable agreement with the Monte Carlo results is obtained for $a=1/4$, $b=1/8$ and $P=0,1$ depending on the parity of the particle number in the Moore-Read liquid. This parity-dependence of the angular momentum serves as an unambiguous demonstration of the non-Abelian nature of the excitations. In addition to the studies of excitations in the Moore-Read liquid, we also apply our scheme to Laughlin liquids, for which we focus on interacting bosonic impurities. With this, the impurities themselves form Laughlin states, which allows for a study of hierarchical fractional quantum Hall states.
\end{abstract}

\maketitle


\section{\label{sec:level1}Introduction}
The most emblematic feature of fractional quantum Hall (FQH) systems are their quasiparticle excitations. Unlike any particle in three dimensions, these quasiparticles are neither bosons nor fermions, but so-called anyons, characterized by intermediate quantum statistics. Importantly, the quasiparticles and their properties also serve for classifying topological phases. In this respect, an important distinction is made between topological phases with Abelian anyons, and the more complex phases, which support non-Abelian anyons. 

In the context of FQH effect, the most prominent Abelian phase occurs at filling factors $\nu=1/m$ (with $m>1$ an odd (for fermions) integer). This phase is well described by the Laughlin wave functions \cite{laughlin83}. Other Abelian FQH phases, occuring at different odd-denominator fillings, can be derived from the Laughlin phase via the so-called hierarchy construction \cite{halperin83,haldane83}. However, there are also FQH phases at even-denominator filling \cite{willett87,shi20}. A prominent wave function to describe a FQH phase in a half-filled Landau level is the Moore-Read wave function, also known as Pfaffian wave function \cite{moore91}. A crucial property of the Moore-Read phase is that it supports non-Abelian excitations, that is, a Moore-Read liquid with $n$ quasiparticles exhibits $2^{n-1}$ degenerate states, and braiding of the quasiparticles is equivalent to a rotation within this degenerate manifold \cite{NAYAK1996529}. Due to the non-commuting nature of different rotations, these excitations are termed ``non-Abelian". This property, together with the manifold's robustness against local noise, has motivated the use of non-Abelian phases for quantum-information processing purposes \cite{nayak08}.

While it is possible to write down models (often called parent Hamiltonians) for which the different FQH wave functions are the exact ground states, it may be hard to determine whether such a phase is indeed realized in a given FQH system. However, at least for the most prominent Abelian phases (such as the $1/3$ Laughlin state), there is no doubt that they are realized in conventional quantum Hall settings (like GaAs quantum wells in strong magnetic fields). In the case of the Laughlin phase, the nature of the FQH phase is well established through theoretical considerations (especially exact numerical studies for small system sizes), but also experimentally, e.g. via transport measurements. Nevertheless, even in this case, a direct experimental evidence for maybe their most important feature, the anyonic excitations, is extremely difficult to obtain. One anyon signature, its fractional charge, has experimentally been determined relatively early via shot noise measurements \cite{saminadayar97}, but attempts to detect also fractional statistics were less successful \cite{lin09,ofek10}, until recent breakthroughs \cite{bartolomei20,nakamura20}.

For the non-Abelian phases, the situation is more controversial. Even from the point of view of exact numerical results, the situation in a half-filled Landau level is much more subtle than at filling 1/3. There is a competition of various different Abelian and non-Abelian phases, and this competition is decided in favor of one or another state by small details in the Hamiltonian \cite{pakrouski15}, such as, for instance, the amount of Landau level mixing. Notwithstanding, a very promising candidate for the half-filled (second) Landau level, i.e. at filling $\nu=5/2$, is the non-Abelian Moore-Read phase \cite{moore91,PhysRevLett.66.3205,READ2001121}. Indeed, also several experimental findings point towards a Pfaffian or anti-Pfaffian phase, including spin polarization \cite{tiemann12}, $e/4$ quasiparticle charge \cite{radu08,dolev08}, or half-integer thermal conductance \cite{banerjee18}. Nevertheles, one must admit also that some experiments suggest an Abelian phase \cite{baer14}. Interferometric measurements to determine the non-Abelian nature of the quasiparticles have not been conclusive \cite{willett09}.

From this perspective, ways to modify or manipulate FQH systems seem desirable, and this goal can be achieved by switching to novel materials or even synthetic quantum systems. For instance, bilayer graphene at half-filled Landau level has shown behavior compatible with the Pfaffian phase \cite{li17,zibrov17}. For single-layer graphene, it has been shown that optical driving can induce non-Abelian topological phases \cite{ghazaryan17,cian20}. An entirely different approach are quantum simulators, which prepare FQH states in highly controlled experimental atomic and/or photonic settings. Advances towards the simulation of FQH physics include the generation of artificial magnetic fields and detection of  topological properties, such as chiral edge states \cite{hafezi13,rechtsman13}, topological quantum numbers \cite{aidelsburger15,mittal16,baboux17,asteria19,tarnowski19}, topological transport \cite{mittal14,bandres16}. Through light-matter coupling, even a Laughlin-like state of two photons has been achieved recently \cite{clark20}, whereas the experimental demonstration of atomic Laughlin states has not yet been conclusive \cite{gemelke10}. In the context of non-Abelian phase engineering, one feature of quantum simulators seems particularly promising: They can operate also with bosonic species, for which often a simple two-body contact potential appears to be sufficient to produce non-Abelian ground states \cite{cooper01,grass12,furukawa12}.

In addition to phase engineering, synthetic quantum Hall systems also provide new detection opportunities: light-matter interactions can be used to create, trap, and braid quasiparticles \cite{paredes01,kapit12,grass14,dutta18,grass18}. The total angular momentum of a FQH system, which for atoms can be measured by time-of-flight imaging, carries signatures of fractional statistics \cite{umucalilar18}. Spectroscopic signatures have been described for atomic systems \cite{cooper14}, graphene \cite{papic18}, or magnetic materials \cite{morampudi17}. Several papers have suggested to bind impurities to fractional quasiparticles \cite{zhang14,zhang15,lundholm16,grusdt16,camacho19,grass20}, which can then be used to trace or manipulate the anyons. 

In the present manuscript, we elaborate on the idea of Ref. \cite{grass20} where the angular momentum of impurities is used to reflect the fractional quantum statistics of Abelian anyons. In the present work, we study whether a similar connection holds for non-Abelian anyons. In this context, Ref. \cite{grass20} has already shown that the angular momentum of a single impurity in a Moore-Read liquid provides a signature of the anyon charge (or the equivalent of charge in atomic systems), and thereby distinguishes between Laughlin-type quasiholes ($\nu/2$ charge) and the true Pfaffian-type elementary excitations ($\nu/4$ charge). In the present paper, we show that also the quantum-statistical behavior of the quasi-holes is reflected by the impurity angular momentum, and their non-Abelian nature is evidenced by a dependence on the parity of the system size.
Specifically, through Monte Carlo sampling of trial wave functions for a Moore-Read liquid with impurities, we obtain for the general form of the quasi-holes' quantum-statistical parameter a functional form $\alpha = a\nu -b +P/2$. For the parameters $a$ and $b$, reasonable numerical agreement is obtained with the values expected for the Moore-Read state ($a=1/4$ and $b=1/8$), and $P=0,1$ depends on the parity of the number of particles of the liquid.

In addition to the study of non-Abelian anyons and their detection via impurity particles, the present paper also generalizes the approach of Ref.~\cite{grass20} in another respect: We demonstrate that non-interacting bosons are not suited as tracer particles, because they form a condensate. However, for repulsively interacting impurities an interesting situation can arise, when the impurities themselves form FQH states. In this case, the impurities can also be bosons, and the scenario allows for exploring the hierarchical construction of FQH states.

Our paper is organized in the following way: In Sec. II, we provide a description of the system and a discussion of the Moore-Read state. In Sec. III, we first give a brief description of the impurity scheme from Ref.~\cite{grass20}, and then present our results for impurities in the Moore-Read liquid. In Sec. IV, we discuss the differences between bosonic and fermionic impurities, and the case of Laughlin-like anticorrelations between the impurities themselves. Finally, in Sec. V, we summarize our main findings and discuss possible continuations of the present work.

\section{Model system and Moore-Read state}
\subsection{System}
We consider two different species of particles confined in the $x-y$ plane by harmonic potentials: majority partcles (a) and minority particles (b). Under a sufficiently strong transverse magnetic field both species are brought in the lowest Landau level, whose basis can be expressed in the symmetric gauge $\mathbf{A}=(B/2)(-y,x,0)$ by Fock-Darwin wave functions
\begin{equation}
\phi_m(z)=(2\pi m!2^m)^{-1/2}z^me^{-|z|^2/4},
\end{equation} 
where $m$ represents the angular momentum of the state. Here the complex coordinate $z=(x+iy)/l_B$ is expressed in terms of the magnetic length $l_B$ that can be set equal for both species.

In presence of repulsive interactions the $a$ particles can form a FQH liquid: we discussed in \cite{grass20} the possibility of a Laughlin liquid, that describes bosonic (fermionic) states at filling $\nu=1/m$ for $m$ even (odd). In the following we will instead consider quantum liquids of spin-polarized electrons well described by the paradigmatic Moore-Read state \cite{moore91}. Differently than in the Laughlin case, the Moore-Read state describes bosons (fermions) for $m$ odd (even). In particular, for $m=2$ this wave function has been proposed to describe the $\nu=5/2$ FQH state for fully spin-polarized electrons \cite{PhysRevLett.66.3205,READ2001121}, when the first Landau level is filled.

\subsection{Moore-Read states} \label{section2a}
The wave function describing the Moore-Read (MR) state for filling $1/m$ is
\begin{equation}
    \psi_{\rm MR}(z)=\text{Pf}\left(\frac{1}{z_i-z_j}\right)\prod_{i<j}(z_i-z_j)^me^{-\sum_i|z|^2/4},
    \label{mooreread}
\end{equation}
where Pf denotes the Pfaffian. The total angular momentum of the state can be read from the polynomial part of the wave function, being equal to the degree of the polynomial in $z_i$. It is given by $L=\frac{m}{2}N_a(N_a-1)-\frac{N_a}{2}$ for $N_a$ particles. The contribution $N_a/2$ is due to the Pfaffian which removes $N_a/2$ zeros from the wave function. 
For $m=1$, the wave function vanishes when three particles are brought in the same point. It is then possible to construct the MR state as the exact ground state of a three-body contact interaction Hamiltonian \cite{GREITER1992567}. Generalizations of this picture are valid for different fillings \cite{PhysRevB.75.195306}.

The simplest zero-energy excitation that this state can host is a  quasi-hole that can be described as in the Laughlin state by multiplying $\psi_{\rm MR}$ by a polynomial term. Explicitly, the wave function reads
\begin{equation}
        \psi_{\rm LQH}(z,w)=\prod_k(z_k-w)\text{Pf}\left(\frac{1}{z_i-z_j}\right)\prod_{i<j}(z_i-z_j)^m,
    \label{holemr}
\end{equation}
where $w$ is the position of the quasi-hole, and we have omitted the exponential factor. The addition of the prefactor implies that the total angular momentum is $L=\frac{m}{2}N_a(N_a-1)+\frac{N_a}{2}$. This quasi-hole, as in the Laughlin case, has fractional charge $e/m$ and Abelian statistical parameter $\alpha=1/m$ \cite{PhysRevB.77.165316}.

More interesting is the fact that in this system each quasi-hole can ``split in two", resulting in a state with the same angular momentum $L$, but with two ``half" quasi-holes (HQH), described by a wave function 
\begin{equation}
    \psi_{\rm HQH}(z,w_1,w_2)=\text{Pf}(W)\prod_{i<j}(z_i-z_j)^m.
    \label{twoimp}
\end{equation}
Here $W$ is a matrix that depends on the parity $P$ of the number of particles $N_a$. If $N_a$ is even, we have
\begin{equation}
    W=\frac{(z_i-w_1)(z_i-w_2)+(i\leftrightarrow j)}{z_i-z_j}.
    \label{matrixw}
\end{equation}
If $N_a$ is odd, this definition would lead to an odd-dimensional matrix, for which the Pfaffian is not defined. Therefore, to obtain $W$ for $N_a$ odd, we have to construct a $N_a+1\times N_a+1$ matrix by adding to the previously defined matrix $W$ a row (column) of 1 (-1), and 0 in the lower right corner \cite{PhysRevLett.123.266801}. 

Similar to the Laughlin case, half quasi-holes are characterized by  fractional charge $e/2m$ and fractional statistics. Crucially, the statistical parameter of the two quasi-holes depends on the parity of $N_a$. In particular, we have that \cite{bonderson11,PhysRevLett.123.266801}
\begin{equation}
    \alpha=\frac{1}{4m}-\frac{1}{8}+\frac{P}{2},
    \label{statparmr}
\end{equation}
where $P=0,1$ for an even (odd) number of particles. 

This expression should be contrasted to the case of a quasihole in the Laughlin liquid. Notably, the statistical parameter $\alpha$ for Pfaffian quasiholes exhibits filling-independent terms, and the $P$-dependence serves as a proof of the non-Abelian statistics of the QHs \cite{PhysRevLett.86.268}. Specifically, the $P$-dependence  reflects the existence of two different fusion channels for the anyons, which, by invoking a conformal field theory description, can be related to the parity of the particle number \cite{bonderson11}. Alternatively, the $P$-dependence can also be explained by the theory of $p$-wave superconductors \cite{PhysRevB.61.10267}. From this viewpoint, the two parity sectors correspond to two degenerate ground states of a $p$-wave superconductor with two half vortices.  The analogy between Pfaffian FQH states and $p$-wave superconductors becomes evident in the composite fermions framework for the state at $\nu=1/2$: in this picture the composite fermions are subjected to a zero effective magnetic field, the state then represents a Fermi liquid that undergoes a BCS instability to a $p$-wave superconducting state.  

An even richer picture appears in the presence of $2n$ HQHs, with $n>1$. The state can still be described by Eq.~\eqref{twoimp}, if we replace $W$ by
\begin{equation}
    \frac{(z_i-w_1)\dots(z_i-w_n)(z_j-w_{n+1})\dots(z_j-w_{2n})+(i\leftrightarrow j)}{z_i-z_j}.
    \label{splitholes}
\end{equation}
It can be seen from Eq.~\eqref{splitholes} that there is an arbitrary choice involved when the $2n$ quasi-holes are split in two groups, each of $n$ elements. This means that it is possible to write more than one such states with $2n$ HQHs. There are $\frac{1}{2}\frac{(2n)!}{n!n!}$ possible ways to group $2n$ elements in two groups, but these states are not orthogonal. Instead, it can be shown that the dimension of the Hilbert space spanned by these degenerate ground states is $2^{n-1}$. Strikingly, an exchange of two quasi-holes can mix one state with one of the others, which is probably the most direct manifestation of the non-Abelian statistics of the HQHs \cite{NAYAK1996529}. 

We also mention the possibility to construct a state that contains a single half quasi-hole by setting
\begin{equation}
    W=\frac{(z_i-w_1)+(z_j-w_1)}{z_i-z_j}.
    \label{wf}
\end{equation}

\noindent The angular momentum of this wave function is $L=\frac{m}{2}N_a(N_a-1)$. In presence of an impurity this state can be retrieved by exact diagonalization of the three-body contact Hamiltonian and a repulsive majority-impurity contact potential \cite{PhysRevB.77.165316}.

\section{Impurities in the Moore-Read liquid}
\subsection{Mean field results}
Before starting the numerical study of the properties of impurities bound to HQHs in the framework of the MR wave function, we can ask what is the effect of the FQH liquid on the single impurity wave function. As we will show here, the binding to the FQH liquid leads to impurity levels characterized through specific average angular momentum values. This observation will later allow us to relate also the anyonic properties of multiple impurities to their total angular momentum.

To describe the effective single-impurity levels, we rely on the assumptions that the impurity is affected by the FQH bath just by a renormalization of the external magnetic field $B$. In particular, from the form of the quasi-hole wave functions, Eqs.~(\ref{holemr}) and (\ref{twoimp}) it follows that the  liquid particle appear as fluxes $\Phi_0$ in the case of Laughlin quasi-holes, or half-fluxes $\Phi_0/2$ in the case of Pfaffian HQHs. Therefore, the net magnetic field becomes $B\rightarrow B(1-\varphi)$ \cite{zhang14}, where $\varphi= \frac{N \Phi_0}{A B}=\nu$ for Laughlin-like quasi-holes, or $\varphi= \frac{N \Phi_0}{2 A B} = \nu/2$ for HQHs, with $A$ being the size of the liquid. As an immediate consequence of the field renormalization, we also get a renormalized length scale $l_b\rightarrow l_b/\sqrt{1-\varphi}$. 

Therefore, an effective wave function for the impurity is given by
\begin{equation}
    \tilde{\phi}_m(w)=\sqrt{\frac{(1-\varphi)^{m+1}}{2\pi m!2^m}}w^m e^{-(1-\varphi)|w|^2/4}.
\end{equation}
By expanding the (squared) amplitude of this wave function in terms of the original wave function amplitudes $|\phi_m(z)|^2$, with $m$ being the angular momentum eigenvalue, we can determine the angular momentum of an impurity in the state $\tilde{\phi}_m$. It is given by
\begin{equation}
    L_m=\frac{m+\varphi}{1-\varphi}.
    \label{singlepartlm}
\end{equation} 
For multiple impurities one might expect to obtain the total angular momentum by filling the available states, so that we obtain $L_f=\sum_{m=0}^{N_b-1}\frac{m+\varphi}{1-\varphi}=\frac{1}{1-\varphi}\left[N_b(N_b-1)/2+N_b\varphi\right]$ for fermionic impurities, or $L_b=N_b\varphi/(1-\varphi)$ for bosonic impurities. However, as we have shown in Ref. \cite{grass20}, impurities bound to quasi-holes in Laughlin liquids become anyonic, and the angular momentum interpolates between $L_f$ and $L_b$:
\begin{equation}
L_{\rm imp}=(1-\alpha)L_f+\alpha L_b.   
\label{prediction}
\end{equation}
Strikingly, the interpolation parameter $\alpha$ is given by the anyonic statistical parameter associated with the exchange of the two quasi-holes. Therefore, by computing $L_f$ and $L_b$ from the effective single-impurity levels, and computing $L_{\rm imp}$ from the many-body wave function, we can determine the statistical parameter of the quasiholes:
\begin{align}
\label{alf}
    \alpha = \frac{L_f-L_{\rm imp}}{L_f-L_b}.
\end{align}
We note that, since $L_{\rm imp}$ reflects the angular momentum of fermionic impurities bound to quasi-holes, a fermionic behavior of these bound states, i.e. $L_{\rm imp}=L_f$, yields a bosonic statistical parameter $\alpha=0$ for the bare quasiholes, whereas bosonic behavior, $L_{\rm imp} =L_b$, corresponds to fermionic quasi-holes, $\alpha=1$.

In the following, we will discuss the validity of this relation for Moore-Read states, where the statistical parameter $\alpha$ of the quasiholes is given by \eqref{statparmr}. With this, our work will provide a possible extension of the detection scheme from Ref. \cite{grass20} to non-Abelian exchange statistics.

\subsection{Numerical results} \label{sec:numres}

If impurities in the lowest Landau level bind to the HQHs of the FQH liquid, the many-body system can still be described by the HQH wave function, Eq.~\ref{twoimp}, but the quasi-hole parameters $w_i$ now become dynamical quantities. In addition, the wave function also has to be multiplied by a factor $e^{-\sum_i\frac{|w_i|^2}{4}}$ for the confinement of the impurities to the lowest Landau level.
In the case of multiple fermionic impurities the state also has to be multiplied by a Vandermonde factor $\prod_{i<j}(w_i-w_j)$ that enforces the Pauli principle. Here, we will restrict ourselves on fermionic impurities. The case of bosonic impurities will be discussed in Sec. IV. 

The resulting wave function is then
\begin{equation}
    \psi(z,w)=\text{Pf}(W)\prod_{k<l,i<j}(w_k-w_l)(z_i-z_j)e^{-\frac{1}{4}\sum_{i,j}|w_i|^2+|z_j|^2},
    \label{simuwf}
\end{equation}
with $W$ chosen appropriately depending on the number of impurities. 

We have used this wave function as a probability distribution for Monte Carlo calculations to compute the expected value of both the impurity angular momentum and, as a crosscheck, the total system angular momenta. Note that while \eqref{simuwf} is not normalized, choosing a Metropolis update rule for the algorithm makes the normalization superfluous.  

\begin{figure}
    \centering
    \includegraphics[width=\linewidth]{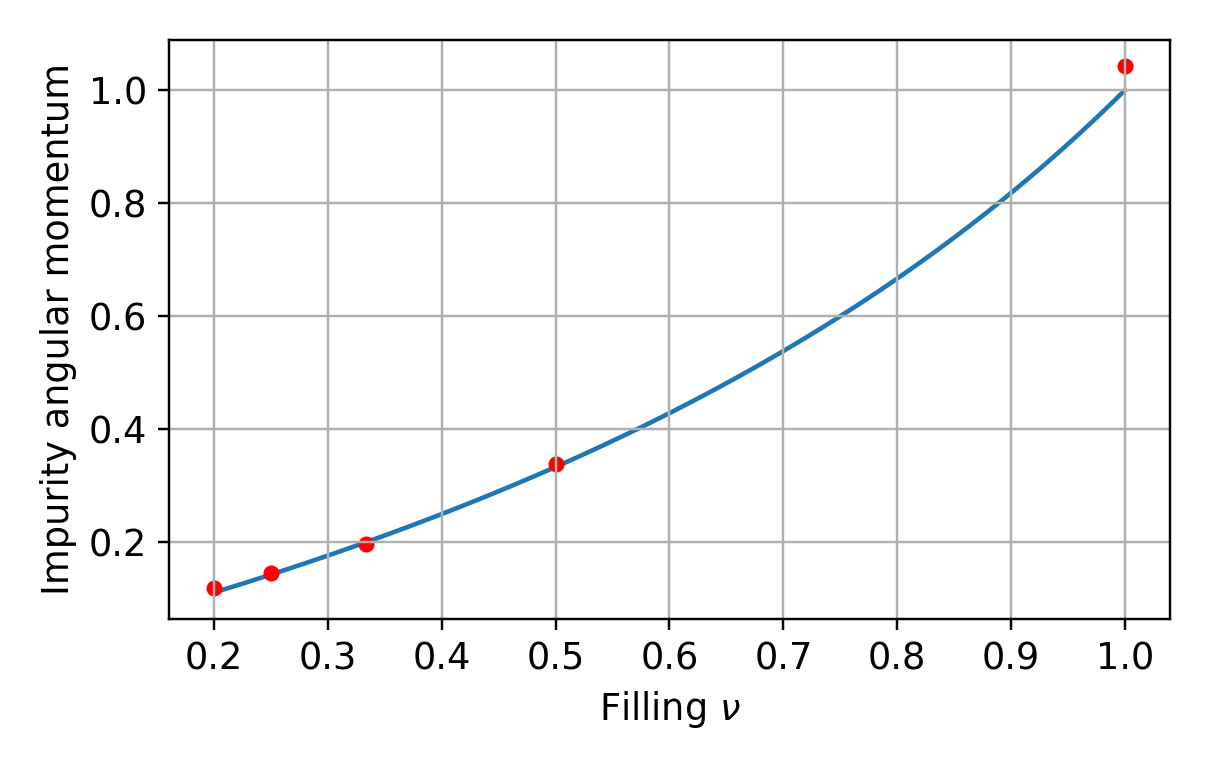}
    \caption{Angular momentum of a single impurity bound to a HQH for states at different fillings (dots), compared to equation \eqref{singlepartlm} for $\varphi=\nu/2$ (solid line). Values obtained by Monte Carlo simulations for 30 majority particles.}
    \label{fig:singlepart}
\end{figure}

First, we have studied the single impurity angular momentum by setting $W$ as in \eqref{wf} for $N_a=30$ with a total angular momentum $L=435/\nu$. The results match well with Eq.~\eqref{singlepartlm} for $m=0$ for a wide range of fillings, as shown in Fig.~\ref{fig:singlepart}. 

\begin{figure}
    \centering
    \includegraphics[width=\linewidth]{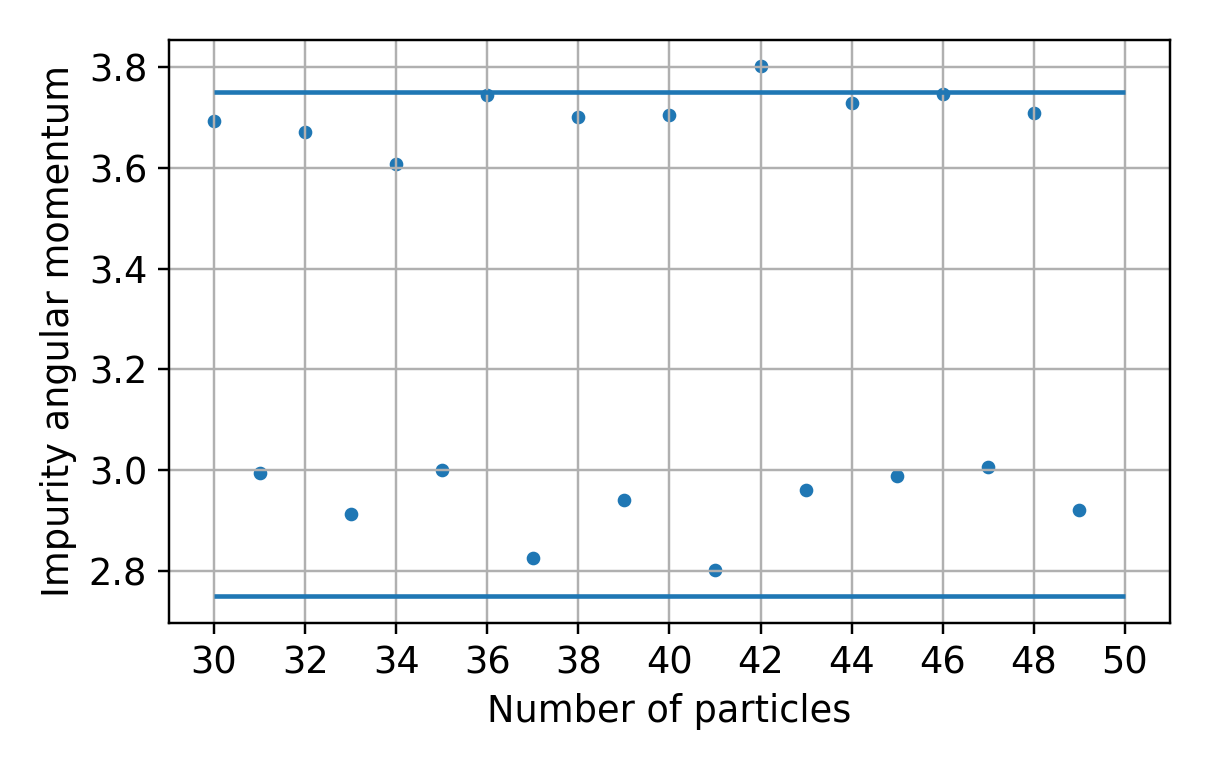}
    \caption{Impurities angular momenta for even/odd number of majority particles at filling $\nu=1$. The jump can be explained considering the different statistical parameter $\alpha$ for the two parity sectors, as described in Eq.~\eqref{statparmr}. The solid lines represent the value of $L_b$ predicted in Eq.~\eqref{prediction} for $P=0,1$. }
    \label{fig:jump}
\end{figure}

Second, we considered the case of two impurity particles, to show that the impurities angular momentum can be used to track the two different parity sectors. Specifically, we computed $L_{\rm imp}$ for $N_a$ from 30 to 49. The expected values of $L_{\rm imp}$ are 3.75 for $N_a$ even and 2.75 for $N_a$ odd.  We show in Fig.~\ref{fig:jump} that for filling $\nu=1$ the jump in angular momentum for even and odd parity is compatible with Eq.~\eqref{prediction}, except for a correction that can be explained by finite size effects, as we will show below.

\begin{figure}
    \centering
    \includegraphics[width=\linewidth]{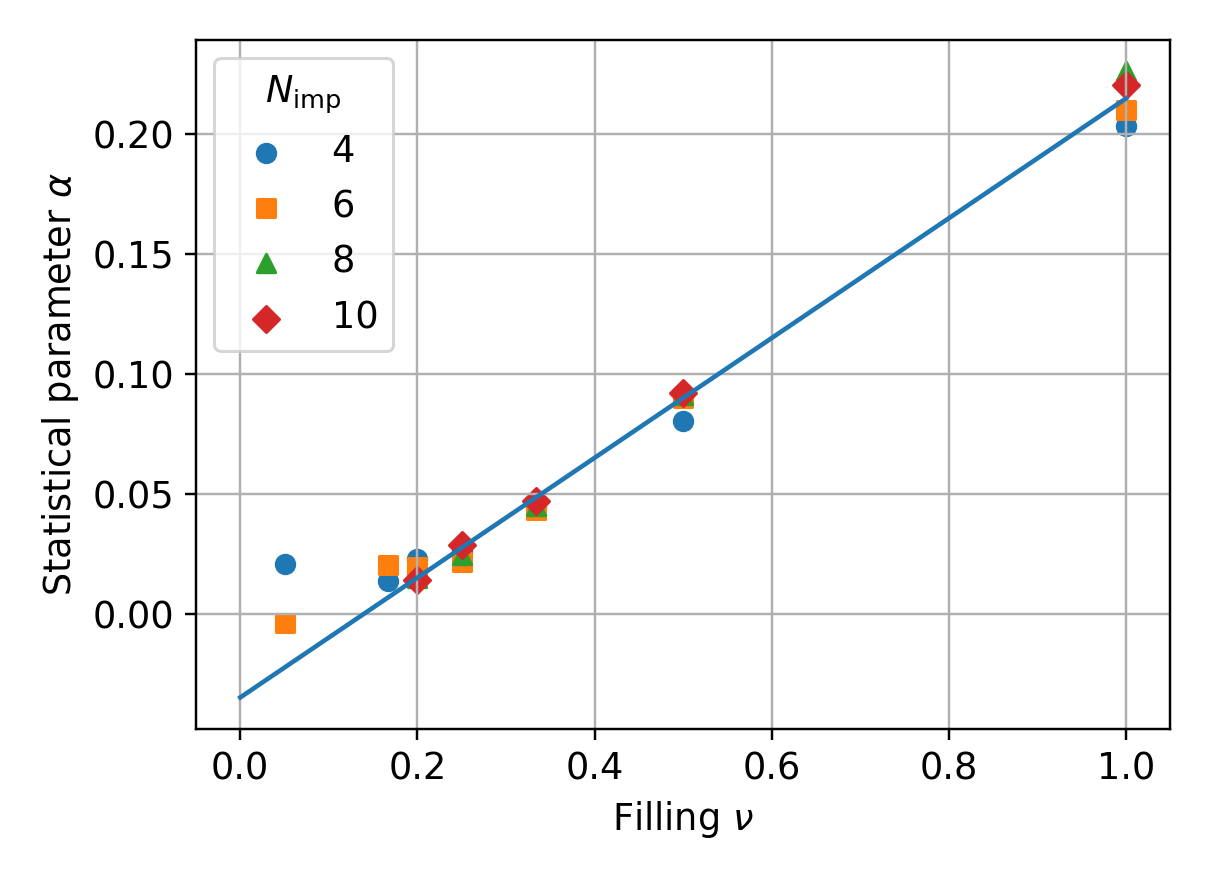}
    \caption{Statistical parameter $\alpha$ of HQHs as a function of filling $\nu$, obtained for different numbers of impurities. For lower fillings (below $1/5$) the HQHs becomes bosonic, and the prediction \eqref{prediction} does not hold, i.e. see the two points at filling $1/20$. Solid line is a fit $\alpha=\nu/4-1/8+0.09$.}
    \label{fig:estrapalpha}
\end{figure}

In order to quantify the statistical parameter $\alpha$ of the anyons, we also studied a larger (even) number of impurities for different filling factors. From each of the numerically computed impurity angular momentum values $L_{\rm imp}$ we extracted the corresponding $\alpha$, via Eq.~\ref{alf}. The results for $N_a=30$ for fillings from 1 to $1/6$ are shown in Fig.~\ref{fig:estrapalpha}, plotting $\alpha$ as a function of filling $\nu$. From the slope of this curve, we see that the filling-dependent part of $\alpha$ perfectly agrees with the expectation, i.e.  $ \alpha \propto \nu/4$. However, the constant contribution is not exactly $-1/8$, as one would expect, but it has a correction of order $10^{-2}$. We account this deviation to the overlapping size of the impurity wave functions. For lower fillings, the prediction \eqref{prediction} breaks down, and the impurities bound to quasi-holes behave effectively as free fermions, and the statistical parameter $\alpha$ of the quasiholes goes to the bosonic limit (zero). 

We also note that the computation of $L_{\rm imp}$ does not lead to any different behaviour for the different $2^{n-1}$ degenerate ground states. Thus, the behavior of the HQHs under braiding cannot be extracted from this impurity angular momentum.

\begin{figure*}
    \centering
    \includegraphics[width=0.75\linewidth]{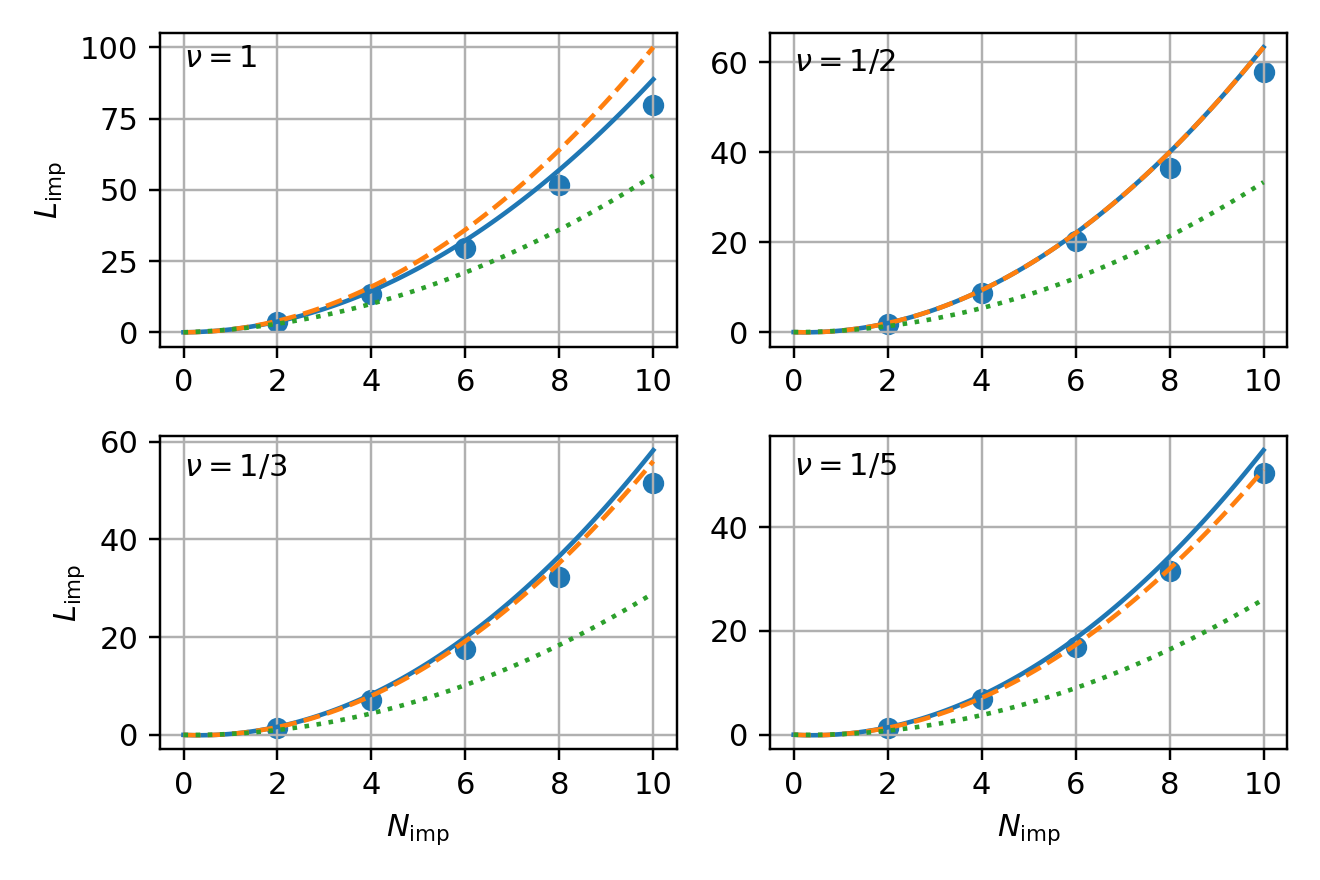}
    \caption{Impurities angular momentum as a function of number of impurities, compared to Eq.~\eqref{prediction} (solid line), value for pure bosonic (dotted line) or fermionic (dashed line) impurities. For low fillings, the impurities behave as free fermions and prediction \eqref{prediction} breaks down. }
    \label{fig:multipart}
\end{figure*}

Our data of $L_{\rm imp}$ for multiple impurities at different $\nu$ is also illustrated in Fig.~\ref{fig:multipart}, plotting $L_{\rm imp}$ vs. the number of impurities in the system at a given $\nu$. We compare this curve with $L_f$ and $L_b$, i.e. with the expectation for fermionic or bosonic particles. At all $\nu \leq 1/2$, the impurity behave very similar to fermions.

\subsection{Analysis of fluctuations of the Berry phase}

Eq.~\eqref{statparmr} holds in a regime where the quasi-holes are sufficently far apart to be considered effectively non-interacting. This in turn influences the validity of Eq.~\eqref{prediction}. For finite distances, the quasi-holes will hybridize and lead to a fluctuation of the exchange statistical phase. These dependence of the statistical parameter on the quasihole distance has been evaluated for MR states in a spherical geometry in \cite{PhysRevLett.90.016802,PhysRevLett.103.076801}. 

In order to estimate this effect in our system, we first need to determine the impurity distance. In our case, this becomes a dynamical variable which we can estimate from the combined wave function. For two fermionic impurities with coordinates $w_1,w_2$ the combined wave function in terms of renormalized Fock-Darwin wave functions is
\begin{equation}
\phi(w_1,w_2)=\frac{\tilde{\phi}_0(w_1)\tilde{\phi}_1(w_2)-\tilde{\phi}_0(w_2)\tilde{\phi}_1(w_2)}{\sqrt{2}},    
\end{equation}
that can be re-expressed in terms of center of mass and relative coordinates $R=(w_1+w_2)/2$, $r=w_1-w_2$ as
\begin{equation}
\tilde{\phi}(r,R)=-\frac{\sqrt{1-\frac{\nu }{2}}(\nu -2)}{8 \pi
   } r
   e^{\frac{(\nu -2)}{16} \left(|r| ^2+4R^2\right)}.
\end{equation}
This function is peaked at $r\sim3$. Monte Carlo computations for the average distance between two impurities in the state \eqref{simuwf} recover roughly the same values for different fillings.

It is then interesting to evaluate directly the statistical phase for two fixed quasi-holes and its dependence on the relative distance. To do so, we considered a configuration with two HQH, one at the center of the system and the other at a fixed radius $R$ from the center. We then compute the Berry phase associated with the state Eq.~\eqref{simuwf} under a rotation of the second hole around the first. To retrieve the statistical phase we have to subtract the Aharonov-Bohm phase that the state accumulates because of the background magnetic field. We compute this phase by performing the same calculations removing the hole placed at the center, describing the system for a single HQH via the state Eq.~(\ref{wf}). Details of the computation can be found in appendix A. 

\begin{figure}
    \centering
    \includegraphics[width=\linewidth]{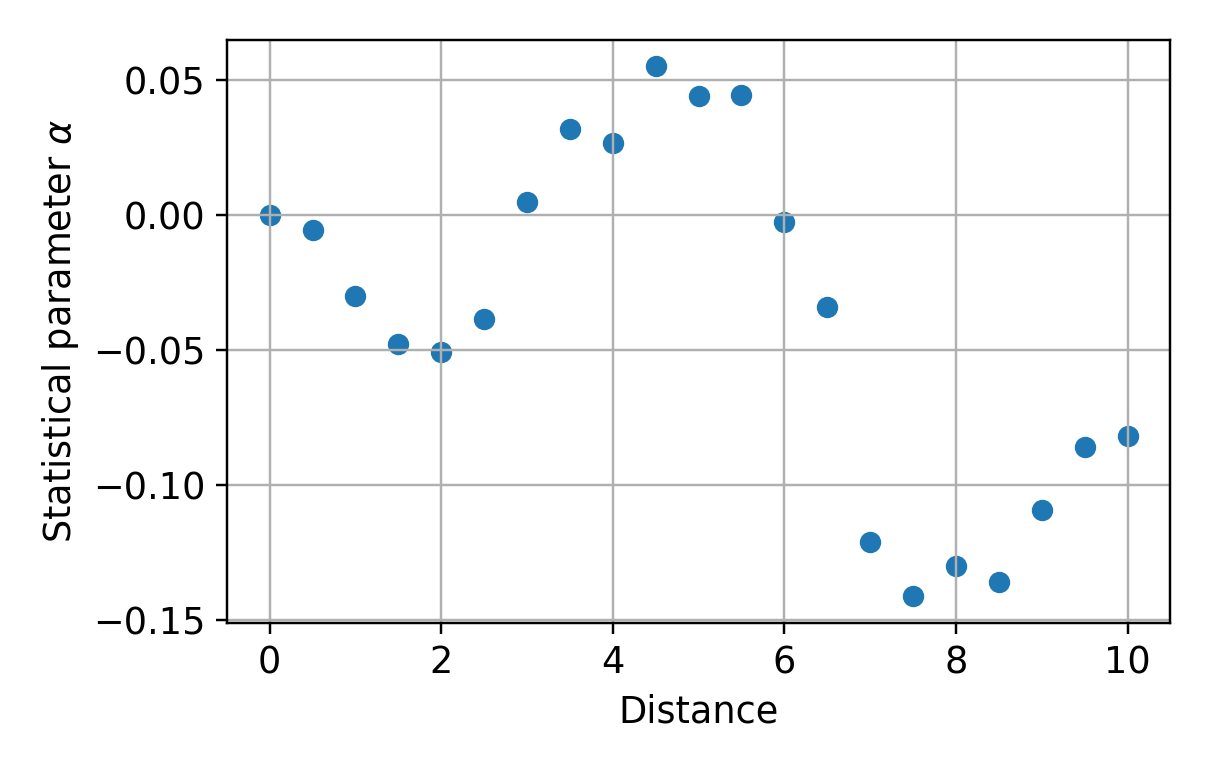}
    \caption{Statistical parameter $\alpha$ computed by explicitly evaluating the Berry phase of a braiding of two quasiholes in a MR state with $\nu=1/2$. For distances of order $r\sim3$ the finite size deviation from the expected value $\alpha=0$ is of order 0.05, compatible with what obtained in Fig.\ref{fig:estrapalpha} }
    \label{fig:berryph}
\end{figure}

As an example, we show the results for the most relevant filling fraction $\nu=1/2$ in Fig. \ref{fig:berryph}. At this filling, one would expect fermionic quasiholes, i.e. impurities with $\alpha=0$. However, at (small) distances, the statistical phase oscillates around zero, with an amplitude on the order of 0.05. This order of magnitude for fluctuations is compatible with the results of section \ref{sec:numres}.

\section{ Anticorrelations of impurities }

All cases we have discussed so far, impurities in Laughlin liquids studied in Ref. \cite{grass20} or impurities in Pfaffian liquids studied in the present work, had one common feature: the impurity particles were always taken as non-interacting fermions, no matter what type of quantum statistics or interparticle interaction governed the liquid of majority particles. In all cases, what accounts for the fermionic nature of the impurities is a Vandermonde factor $\prod_{i<j} (w_i-w_j)$ which multiplies the quasihole wave function, cf. Eq.~\eqref{simuwf}. The Vandermonde factor produces the minimum anticorrelations required for fermionic impurities by the Pauli principle, but one may ask what would happen if these anticorrelations were lacking in the total wave function - a scenario which would be valid for non-interacting bosonic impurities? Or, contrarily, one could also imagine systems, with either fermionic or bosonic impurities, in which interactions between the impurity particles give rise to stronger anticorrelations beyond the one from the Vandermonde determinant. In the present section, we will study these cases, and we will show how anticorrelations between the impurities reflect in the total impurity angular momentum. For the sake of simplicity, we will restrict our considerations in this section to the case of impurities in an Abelian Laughlin liquid.

\subsection{Non-interacting bosonic impurities}
Let us consider a Laughlin liquid at filling $\nu=1/q$, described by a wave function $\Psi_q$. Quasiholes in this liquid, at positions $w_i$, are described by the wave function $ \Psi(\{w_i\}) = \prod_{i,j} (w_i-z_j) \Psi_q$, where $i$ runs over all quasiholes and $j$ over all liquid particles. 
Multiplying this wave function by the Vandermonde term $\prod_{i<j} (w_i-w_j)$, accounts for the binding of quasiholes to non-interacting fermions, and these fermions carry an average angular momentum, $L_{\rm imp}$, which interpolates between the values for free fermions and free bosons, cf. Eq.~\eqref{prediction}. This implies that $L_{\rm imp}$ scales extensively with the number of impurities.

In contrast, the appropriate wave function for non-interacting bosonic impurities bound to quasiholes is simply given by $ \Psi(\{w_i\}) = \prod_{i,j} (w_i-z_j) \Psi_q$, without the Vandermonde determinant. The values for $L_{\rm imp}$, found by Monte Carlo sampling of this wave function, are given in Table \eqref{tabbose} for some values of $q$, $N$ (number of particles in the liquid), and $N_{\rm imp}$ (number of impurities), and contrasted to the analog values obtained in the case of fermionic impurities. Interestingly, for a given $q$, the value obtained for the bosonic impurities is approximately constant, i.e. it depends neither on the number of particles in the liquid (which is true also in the case of fermionic impurities), nor on the number of impurities (in stark contrast to the case of fermionic impurities). The value of total angular momentum for the bosonic impurities appears to be proportional to the average angular momentum of a single impurity in its ground state, $L_0=\frac{1}{q-1}$, as given by Eq.~\ref{singlepartlm}. In fact, for all cases shown in Table \ref{tabbose}, we approximately have $L_{\rm imp} \approx 1.4 L_0$.
\begin{table}
\begin{tabular}{c|c|c|c|c}
$q$ & $N_{\rm imp}$ & $N$ & $L_{\rm imp}$ (bosonic) & $L_{\rm imp}$ (fermionic) \\
\hline
\hline
3 & 2 & 10 & 0.70 & 2.03 \\
3 & 3 & 10 & 0.73 & 4.55 \\
3 & 4 & 10 & 0.72 & 8.11 \\
\hline
5 & 2 & 7  & 0.33 & 1.47 \\
5 & 3 & 7  & 0.32 & 3.65 \\
\hline
2 & 2 & 7  & 1.39 & 3.11 \\
2 & 3 & 7  & 1.43 & 6.31
\end{tabular}
\caption{ \label{tabbose} For different filling $\nu=1/q$ of a Laughlin liquid with $N$ particles, we numerically obtain the average angular momentum $L_{\rm imp}$ of $N_{\rm imp}$ non-interacting impurities, which can be either bosonic or fermionic.}
\end{table}

An explanation for this curious behavior could be the following: The bosonic impurities form a condensate (in which the individual impurities fluctuate around the condensate center of mass), and all quasiholes bind to this condensate, just as if there was only a single quasihole and a single impurity. In this picture, the (small) difference between $L_{\rm imp}$ and $L_0$ would then be due to the fluctuations of impurities within the impurity condensate, although the picture does not necessarily imply that $L_{\rm imp}$ is independent from $N_{\rm imp}$.

What appears to be clear, though, is the fact that non-interacting bosons as impurity particles are not suited for probing the anyonic properties of quasiholes. In the following subsection, we are going to investigate whether and how the situation changes if the bosonic impurities are interacting.

\subsection{Interacting impurities}
We restrict our study of interacting impurities to the simplest and most relevant case of bosonic impurities with repulsive contact interaction. With the impurities being subject to Landau quantization, this implies that the zero-energy ground state of the impurities itself is a bosonic Laughlin state, $\sim \prod_{i<j} (w_i-w_j)^2$. Therefore, in this case, the Vandermonde determinant in the wave function of fermionic impurities, has to be replaced by these bosonic Laughlin-like anticorrelations. Evaluating again the average impurity angular momentum value numerically, we find that it matches very well with the following pattern:
\begin{align}
L_{\rm imp}(N_{\rm imp},q) = \frac{N_{\rm imp} + \frac{2q-1}{2}N_{\rm imp}(N_{\rm imp}-1)}{q-1}.
\label{pattern}
\end{align}
This observation immediately leads to the question how this expression connects to the effective single-impurity levels $L_m$, given by Eq.~\eqref{singlepartlm}. A simple guess would be that each impurity pairs enters a state in which their relative angular momentum is given by $L_2$, as the impurities are forming a $q=2$ Laughlin state. This guess, though, does not match with the observed pattern. However, two more things should be considered: (i) Apart from the relative angular momentum of pairs, the Laughlin liquid also has a center-of-mass angular momentum which comes from  $N_{\rm imp}$ particles condensing into $L_0$. Thus, $L_{\rm com} = N_{\rm imp}/(q-1).$ (ii) For the relative angular momentum, one has to consider screening effects, because in the vicinity of one impurity/quasihole the majority density is lowered. Following the arguments of Ref. \cite{zhang14}, we first note that a wave function of a pair at relative angular momentum $M$ has an amplitude peak at radial distance $R_M = (2M)^{1/2} l_B$ in the absence of any screening. The screening due to the majority liquid effectively leads to a redefinition of the magnetic length, $l_B \rightarrow l_B^* = l_B/\sqrt{1-\nu}$. The screening which one impurity experiences due to the presence of the other impurity is captured by $M \rightarrow M^* = M-\nu$. Thus, for a pair at $M=2$, the effective size of the wave function is given by $R^* = [2(M-\nu)]^{1/2}l_B^* = \left[ \frac{2(M-\nu) }{1-\nu} \right]^{1/2} l_B$. The corresponding effective relative angular momentum is $L^* = \frac{1}{2} (R^*/l_B)^2 = \frac{M-\nu}{1-\nu} = \frac{Mq-1}{q-1}$. Thus, for $N_{\rm imp}(N_{\rm imp}-1)/2$ pairs at $M=2$, the relative angular momentum becomes in total:
\begin{align}
L_{\rm rel} = \frac{1}{2} N_{\rm imp}(N_{\rm imp}-1)\frac{2q-1}{q-1}
\end{align}
With this, the sum, $L_{\rm rel}+L_{\rm com}$ exactly matches the pattern in Eq.~\eqref{pattern} found numerically.

Another way of understanding Eq.~(\ref{pattern}) is in the light of hierarchy states \cite{halperin83,haldane83}. This construction builds upon the Laughlin states at filling $1/q$ (or their hole-conjugate at filling $1-1/q$). It then argues that fractional quantum Hall states at other (odd-denominator) filling factors can appear when quasiholes or quasiparticles in the parent liquid themselves form a Laughlin-like state. Noting the relation between filling factor $\nu$, angular momentum $L$, and particle number $N$, $\nu = {\rm lim}_{N\rightarrow \infty} \frac{N^2}{2L}$ \cite{macdonald85}, we find that the angular momentum of Eq.~(\ref{pattern}) corresponds to a fractional quantum Hall state at $\nu=\frac{q-1}{2q-1}$, which matches the filling factor of the first state in the hierarchical construction. This observation suggests a feasible way of exploring hierarchical fractional quantum Hall states using bosonic impurities with repulsive contact interactions.

\section{Summary and Outlook}

In this work, we study the angular momentum of non-interacting fermionic impurities, bound to quasiholes in FQH states constrained to lowest Landau levels, and represented by Moore-Read wave functions, to determine the non-Abelian statistics of fractional quasiparticle excitations. Such FQH states can host Laughlin like quasi-holes as zero energy excitations, with fractional charge and Abelian statistics. However, interestingly the Laughlin like quasi-holes can also split into states that can host HQHs with non-Abelian statistics, described by Pfaffian wave functions.

When dynamical impurities bind to such HQHs, we show that the quantum-statistical behavior of these objects can be directly read from the impurity angular momentum. The impurity particles see a renormalized magnetic field in the presence of majority particles, which in turn depends on the charge of the HQH. It should be possible to determine the angular momentum of many impurities by taking into account the renormalization of the magnetic field for a single impurity, and by filling the single impurity angular momentum levels, assuming them to be either fermions or bosons. Remarkably, we see here (and in our Ref.~\cite{grass20}), via Monte Carlo sampling of the many-body wave function, that the angular momentum of many impurities does not quench either of the limits, bosonic or fermionic, but actually interpolates between the two, capturing the anyonic statistics. The interpolation parameter for impurities bound to HQHs is filling dependent. Its slope and a parity-dependent intercept are the same as the statistical parameters we expect from the Moore-Read state. It is in fact the parity of the particle number that truly reflects the non-Abelian nature here, in contrast to the Abelian behavior seen in case of Laughlin liquids. The statistical parameter extracted from the angular momentum of impurities reasonably captures the change in intercept due to parity, confirming the non-Abelian nature of the anyons. There are, however, some fluctuations of the intercept, due to finite size effects, that can be well understood by studying the Berry phase in a finite system, upon exchange of impurities.

Moreover, instead of non-interacting fermionic impurities detailed above, we also look at non-interacting bosonic impurities as tracer particles, but in the much simpler Abelian situation. Through Monte Carlo simulations, we demonstrate that total angular momenta for such bosonic impurities appear to be proportional to the average angular momentum of a single impurity in its ground state. This happens because, confined to the lowest Landau level, such bosons form a condensate. Therefore, we clearly see that such bosonic impurities are not suitable as tracer particles to extract the statistical behaviour. However, if the bosons can repulsively interact with each other, the situation becomes much more intriguing. We then numerically calculate the total angular momentum of such impurities and find that they are appropriately explained by considering the total center of mass angular momentum and the screened (due to other impurities) relative angular momentum. Intriguingly, the total angular momentum corresponds to a FQH state which matches the filling factor of the first state in the hierarchical construction of odd denominator FQH states. Such odd denominator states arise from quasi-particles in the parent Laughlin liquid, forming their own FQH states.

Therefore, our study not only opens up the experimentally challenging possibility of directly reading the non-Abelian statistics of the Moore-Read states via measurement of the impurity angular momentum (equivalent to impurity density measurements), it also shows how other odd denominator Laughlin states, understood in terms of the hierarchical construction, can be probed within the same approach.

\acknowledgements{
ICFO group acknowledges support from ERC AdG NOQIA, Agencia Estatal de Investigación (“Severo Ochoa” Center of Excellence CEX2019-000910-S, Plan National FIDEUA PID2019-106901GB-I00/10.13039 / 501100011033, FPI), Fundació Privada Cellex, Fundació Mir-Puig, and from Generalitat de Catalunya (AGAUR Grant No. 2017 SGR 1341, CERCA program, QuantumCAT $\_$U16-011424 , co-funded by ERDF Operational Program of Catalonia 2014-2020), MINECO-EU QUANTERA MAQS (funded by State Research Agency (AEI) PCI2019-111828-2 / 10.13039/501100011033), EU Horizon 2020 FET-OPEN OPTOLogic (Grant No 899794), and the National Science Centre, Poland-Symfonia Grant No. 2016/20/W/ST4/00314.
  N.B. acknowledges support from a ``la Caixa” Foundation (ID 100010434) fellowship. The fellowship code is  LCF/BQ/DI20/11780033. B. J.-D. acknowledges funding from Ministerio de Economia y Competitividad Grant No FIS2017-87534-P.
  U.B. acknowledge
support by the “Cellex-ICFO-MPQ Research Fellows”, a
joint program between ICFO and MPQ - Max-Planck Institute for Quantum Optics, funded by the Fundació
Cellex.
  T.G. acknowledges financial support from a fellowship granted by “la Caixa” Foundation (ID 100010434, fellowship code LCF/BQ/PI19/11690013).}

\appendix
\section{Berry phase computation}
We are interested in computing the Berry phase associated to a braiding of a HQH around another fixed in the center of the system. The system is then described by Eq.~\eqref{twoimp} with $w_1=0$, $w_2=Re^{i\theta}$. The Berry phase for an adiabatic change of the parameter $\theta$ at fixed $R$ is then
\begin{equation}
    \gamma=i\oint\text{d}\theta\langle\frac{\psi^*}{\mathcal{N}}\frac{\text{d}}{\text{d}\theta}\frac{\psi}{\mathcal{N}}\rangle,
\end{equation}
where $\mathcal{N}=\sqrt{\langle\psi^*\psi\rangle}$ is the normalization of the wave function. As this normalization is explicitly dependent on the value of $\theta$ we have to explicitly consider it in the computation of the derivative obtaining
\begin{equation}
    \gamma=i\oint\text{d} \theta\,\text{Im}\left(\frac{1}{\mathcal{N}^2}\langle\psi^*\frac{\text{d}}{\text{d}\theta}\psi\rangle\right)=i2\pi\text{Im}\left(\frac{1}{\mathcal{N}^2}\langle\psi^*\frac{\text{d}}{\text{d}\theta}\psi\rangle\right),
\end{equation}
where the last step is justified by the rotational invariance of the system.

There is left to evaluate the derivative in the round brackets.
The only dependence on $\theta$ in the state \eqref{twoimp} is in the Pfaffian term. We then can use the identities
\begin{eqnarray}
\frac{\text{d}}{\text{d}\theta}\text{det}W=\frac{\text{d}}{\text{d}\theta}[\text{Pf}(W)]^2,\\
\frac{\text{d}}{\text{d}\theta}\text{det}(W)=\text{det}(W)\text{Tr}\left(W^{-1}\frac{\text{d}W}{\text{d}\theta}\right),
\end{eqnarray}
to show that 
\begin{equation}
\frac{\text{d}\psi}{\text{d}\theta}=\frac{1}{2}\text{Tr}\left(W^{-1}\frac{\text{d}W}{\text{d}\theta}\right)\psi.
\end{equation}
We then obtain 
\begin{equation}
    \gamma=i\pi\text{Im}\left(\frac{1}{\mathcal{N}^2}\left\langle\psi^*\text{Tr}\left(W^{-1}\frac{\text{d}W}{\text{d}\theta}\right)\psi\right\rangle\right).
    \label{gamma}
\end{equation}

The quantity inside the round bracket can then be computed by Monte Carlo. To extract the statistical phase we can remove the Aharonov-Bohm contribution by removing the $w_1$ quasi-hole: this corresponds to substituting $W,\psi$ in Eq.~\eqref{gamma} with the appropriate ones for a single HQH. Finally the statistical parameter is $\alpha=\gamma/(2\pi)$.

\bibliography{biblio}

\end{document}